\documentclass[prd,amsmath,amssymb,showpacs,superscriptaddress,nofootinbib,twocolumn]{revtex4-2}
\usepackage{multirow}
\usepackage{float}
\usepackage{graphicx}
\usepackage{dcolumn}
\usepackage{bm}
\usepackage{rotating}
\usepackage{color}
\usepackage{subfigure}
\usepackage{pstricks}
\usepackage{soul}
\usepackage{overpic}
\usepackage[english]{babel}
\usepackage[colorlinks,linkcolor=blue,anchorcolor=blue,citecolor=blue]{hyperref}
\usepackage{tabularx}
\usepackage{ulem}
\usepackage{lineno}

\def \reJ {(6.34 \pm 0.21 \pm 0.37)\times 10^{-5}}
\def \reP {(9.59 \pm 2.37 \pm 0.61)\times 10^{-6}}
\def \reJns {(6.34 \pm 0.21)\times 10^{-5}}
\def \rePns {(9.59 \pm 2.37)\times 10^{-6}}
\def \reQ {15.1 \pm 3.8}

\begin{document}

\title{\boldmath Observation of the $J/\psi$ and $\psi(3686)$ decays into $\eta\Sigma^{+}\overline{\Sigma}{}^-$}
\author{\small
M.~Ablikim$^{1}$, M.~N.~Achasov$^{12,b}$, P.~Adlarson$^{72}$, M.~Albrecht$^{4}$, R.~Aliberti$^{33}$, A.~Amoroso$^{71A,71C}$, M.~R.~An$^{37}$, Q.~An$^{68,55}$, Y.~Bai$^{54}$, O.~Bakina$^{34}$
% ...
\\
\vspace{0.2cm}
(BESIII Collaboration)\\
\vspace{0.2cm} {\it
$^{1}$ Institute of High Energy Physics, Beijing 100049,  People's Republic of China\\
% ...
$^{4}$ Bochum  Ruhr-University, D-44780 Bochum, Germany\\
% ...
$^{12}$ G.I. Budker Institute of Nuclear Physics SB RAS (BINP), Novosibirsk 630090, Russia\\
% ...
$^{33}$ Johannes Gutenberg University of Mainz, Johann-Joachim-Becher-Weg 45, D-55099 Mainz, Germany\\
$^{34}$ Joint Institute for Nuclear Research, 141980 Dubna, Moscow region, Russia\\
% ...
$^{37}$ Liaoning Normal University, Dalian 116029, People's Republic of China\\
% ...
$^{54}$ Southeast University, Nanjing 211100, People's Republic of China\\
$^{55}$ State Key Laboratory of Particle Detection and Electronics, Beijing 100049, Hefei 230026, People's Republic of China\\
% ...
$^{68}$ University of Science and Technology of China, Hefei 230026, People's Republic of China\\
% ...
$^{71}$ University of Turin and INFN, (A)University of Turin, I-10125, Turin, Italy; (B)University of Eastern Piedmont, I-15121, Alessandria, Italy; (C)INFN, I-10125, Turin, Italy\\
$^{72}$ Uppsala University, Box 516, SE-75120 Uppsala, Sweden\\
% ...
$^{b}$ Also at the Novosibirsk State University, Novosibirsk, 630090, Russia.\\
% ...
}\vspace{0.4cm}}


\begin{abstract}
The decays $J/\psi\to\eta\Sigma^{+}\overline{\Sigma}{}^-$ and $\psi(3686)\to\eta\Sigma^{+}\overline{\Sigma}{}^-$ are observed for the first time, using $(10087 \pm 44)\times 10^{6}$ $J/\psi$ and $(448.1 \pm 2.9)\times 10^{6}$ $\psi(3686)$ events collected with the BESIII detector at the BEPCII collider. We determine the branching fractions of these two decays to be ${\cal B}(J/\psi\to\eta\Sigma^{+}\overline{\Sigma}{}^-)=\reJ$ and ${\cal B}(\psi(3686)\to\eta\Sigma^{+}\overline{\Sigma}{}^-)=\reP$, where the first uncertainties are statistical and the second are systematic. The ratio of these two branching fractions is determined to be $\frac{{\cal B}(\psi(3686)\to\eta\Sigma^{+}\overline{\Sigma}{}^-)}{{\cal B}(J/\psi\to\eta\Sigma^{+}\overline{\Sigma}{}^-)}=(\reQ)\%$, which is in agreement with the ``12\% rule."
\end{abstract}

\maketitle

\section{\boldmath Introduction}

Studies of the hadronic decays of the $c\bar{c}$ states $J/\psi$ and $\psi(3686)$ (here referred to as $\psi$) provide good opportunities to test theories in the transition region of perturbative and nonperturbative quantum chromodynamics (QCD), as well as valuable information on the structure of charmonia~\cite{intro0}.

Many kinds of two-body decays of charmonia into a baryon pair, i.e. $\psi\to B\bar{B}$ ($B$ stands for a baryon), have been observed in experiments, and they have been understood in terms of $c\bar{c}$ annihilations into three gluons or into a virtual photon~\cite{intro1}. The measurement of three-body decays $\psi\to B\bar{B}P$, where $P$ stands for a pseudoscalar meson such as $\eta$ or $\pi^0$, has the additional advantage to study the intermediate excited hadrons.
On this field, so far the BESIII Collaboration has published the studies on the decays $\psi\to p\bar{p}\pi^{0}(\eta)$~\cite{eeetall2} and $\psi\to\Lambda\bar{\Lambda}\pi^{0}(\eta)$~\cite{bes3etallwang}, while the similar isospin-allowed decay $\psi\to\eta\Sigma^{+}\overline{\Sigma}{}^-$ has not yet been measured. In addition, since most of the excitation spectra of hyperons are still not well understood~\cite{intro2}, the $\psi\to\eta\Sigma^{+}\overline{\Sigma}{}^-$ decay provides a good opportunity to search for potential $\Sigma$ excitations.

Perturbative QCD (pQCD) predicts that the ratio between the branching fractions of $J/\psi$ and $\psi(3686)$ decaying into the same final states obeys the so-called ``12\% rule"~\cite{12rule1,12rule2}, expressed by $\frac{{\cal B}(\psi(3686)\to X)} {{\cal B}(J/\psi\to X)}\approx12\%$, where $X$ denotes any exclusive hadronic decay mode or the $\l^{+}\l^{-}~(\l=e,~\mu)$ final state. A large fraction of measured branching fractions for exclusive decays follows the ``12\% rule" within errors. However, the measured ratio of ${\cal B}(\psi(3686)\to\rho\pi)$ to ${\cal B}(J/\psi\to\rho\pi)$ is much less than the prediction. To understand the deviation from ``12\% rule" in some decay modes, many theoretical and experimental efforts have been made. For example, the ratio for the isospin violating decay $\psi\to\Lambda\bar{\Lambda}\pi^0$ deviates from 12\%, while it is consistent for the isospin-allowed decay $\psi\to\Lambda\bar{\Lambda}\eta$~\cite{bes3etallwang}. The BESIII experiment has collected the largest data sample of $J/\psi$ and $\psi(3686)$ events, providing a good opportunity to test the ``12\% rule" in the decays involving $\Sigma$ hyperons in the final state.

In this paper, we report the first measurements of the branching fractions of $J/\psi\to\eta\Sigma^{+}\overline{\Sigma}{}^-$ and $\psi(3686)\to\eta\Sigma^{+}\overline{\Sigma}{}^-$, based on the data samples of $(10087 \pm 44)\times 10^{6}$ $J/\psi$ events and $(448.1 \pm 2.9)\times 10^{6}$ $\psi(3686)$ events~\cite{Njpsi,Npsip} collected with the BESIII detector. Besides, we search for potential excited baryon states and unknown structures in the $\eta\Sigma$ and $\Sigma^{+}\overline{\Sigma}{}^-$ invariant mass spectra.

\section{\boldmath BESIII Detector and Monte Carlo Simulation}
The BESIII detector is a magnetic spectrometer~\cite{BESIII} located at the electron positron collider BEPCII. The cylindrical core of the BESIII detector consists of a helium-based multilayer drift chamber (MDC), a plastic scintillator time-of-flight system (TOF), and a CsI (Tl) electromagnetic calorimeter (EMC), which are all enclosed in a superconducting solenoidal magnet providing a 1.0 T (0.9 T in 2012) magnetic field. The solenoid is supported by an octagonal flux-return yoke with resistive plate counter muon identifier modules interleaved with steel. The acceptance of charged particles and photons is 93\% over 4$\pi$ solid angle. The charged-particle momentum resolution at 1 GeV/$c$ is 0.5\%, and the specific ionization energy loss ($dE/dx$) resolution is 6\% for the electrons from Bhabha scattering. The EMC measures photon energies with a resolution of 2.5\% (5\%) at 1 GeV in the barrel (end cap) region. The time resolution of the TOF barrel part is 68 ps, while that of the end cap part is 110 ps. The end cap TOF system was upgraded in 2015 with multigap resistive plate chamber technology, providing a time resolution of 60 ps~\cite{BESTOF1,BESTOF2}.

To determine the reconstruction efficiency of the decay channels, exclusive MC samples are simulated by using the phase space (PHSP) model for the decay of each reaction channel. These samples are produced with a {\sc GEANT4}-based~\cite{BESIII5} Monte Carlo (MC) package, which includes the geometric description of the BESIII detector and the detector response. The simulation also models the beam energy spread and initial state radiation (ISR) in the $e^+ e^-$ annihilations with the generator {\sc KKMC}~\cite{BESIII6}.
For the determination of background contributions, the so-called inclusive MC samples are used. These samples include the production of the $J/\psi$ or $\psi(3686)$ events as resonance, in ISR production of the $\psi$, and as continuum processes as incorporated in {\sc KKMC}. For these decays all known modes are modeled with {\sc EVTGEN}~\cite{BESIII7,BESIII7p2} using branching fractions taken from the Particle Data Group (PDG)~\cite{PDG}. All remaining unknown decays of charmonium states are modeled with {\sc LUNDCHARM}~\cite{BESIII9}. Final state radiation from charged final state particles is incorporated using {\sc PHOTOS}~\cite{BESIII10}.

\section{\boldmath Event selection}
In the channel $\psi\to\eta\Sigma^{+}\overline{\Sigma}{}^-$, the $\Sigma^{+}$($\overline{\Sigma}^{-}$) is reconstructed with $\Sigma^+\to p\pi^{0}(\overline{\Sigma}^{-}\to\bar{p}\pi^{0})$, while the $\pi^{0}$ and $\eta$ are reconstructed with $\pi^{0}\to\gamma\gamma$ and $\eta\to\gamma\gamma$, respectively.

For each charged track, the distance of closest approach to the interaction point (IP) is required to be within 20 cm along the beam direction, while no requirement in the plane perpendicular to the beam direction is applied. Charged tracks detected in the MDC are required to be within a polar angle ($\theta$) range of $|\cos\theta|<0.93$, where $\theta$ is defined with respect to the $z$ axis. The measurements of the flight time in the TOF and of the $dE/dx$ in the MDC are combined to compute particle identification (PID) confidence levels for pion, kaon and proton hypotheses. The track is assigned to the particle type with the highest confidence level. One proton and one antiproton are required to be identified.

Photon candidates are reconstructed from isolated showers in the EMC within 700 ns from the event start time. Their energy is required to be greater than 25 MeV in the barrel region ($|\cos\theta|<0.8$) and 50 MeV in the end cap region ($0.86<|\cos\theta|<0.92$). The $\pi^{0}$ and $\eta$ candidates are selected from all the photon pairs by a selection on invariant mass of (0.110, 0.160) and (0.450, 0.650) GeV/$c^{2}$, respectively. Furthermore, events are required to contain at least one $\eta$ and two $\pi^0$ candidates.

In order to suppress the remaining backgrounds and to improve the mass resolution, a seven-constraint (7C) kinematic fit is performed on the $\eta\pi^0\pi^0 p\bar{p}$ candidates, by constraining  the total four-momentum of the final state particles to the total initial four-momentum of the colliding beams, and the invariant mass of the two photons from the decay of the $\eta/\pi^0$ to the nominal mass value.
If there is more than one combination surviving the selections, the one with the least $\chi^{2}_{\rm 7C}$ of the kinematic fit is selected. Furthermore, the $\chi^{2}_{\rm 7C}$ value is required to be less than 30 and 25 for $J/\psi$ and $\psi(3686)$ decays, respectively, by optimizing the figure of merit (FOM), defined as $S/\sqrt{S+B}$, where $S$ is the number of signal events from the signal MC sample and $B$ is the number of background events from the inclusive MC sample.
Since the masses of the $\Sigma^{+}$ and $\overline{\Sigma}^{-}$ candidates are not constrained in the fit, the two $\pi^0$ from $\Sigma^{+}$ and $\overline{\Sigma}^{-}$ decays are selected by minimizing $\Delta=\sqrt{(M_{p\pi^0}-m_{\Sigma^{+}})^2 + (M_{\bar{p}\pi^{0} }-m_{\overline{\Sigma}^{-}})^2}$ by iterating all the possible proton/antiproton and $\pi^0$ combinations.

For $J/\psi\to\eta\Sigma^{+}\overline{\Sigma}{}^-$, the background from $J/\psi\to p\bar{p}\eta'$ is vetoed by requiring the invariant mass of the $\eta\pi^0\pi^0$ combination to be outside the $\eta'$ signal region [0.95, 0.97] GeV/$c^{2}$.
For $\psi(3686)\to\eta\Sigma^{+}\overline{\Sigma}{}^-$, the recoil mass of the $\eta$ is required to satisfy $M_{\eta}^{\rm rec}<3.050$ GeV/c$^{2}$ to suppress the backgrounds from $\psi(3686)\to\eta J/\psi$ and $\psi(3686)\to\gamma\chi_{c0,1,2},~\chi_{c0,1,2}\to\gamma J/\psi$ with $J/\psi\to\Sigma^{+}\overline{\Sigma}{}^-$. The background from $\psi(3686)\to\pi^{0}\pi^{0}J/\psi$ is vetoed by requiring the recoil mass of the $\pi^{0}\pi^{0}$ pair to be outside the $J/\psi$ signal region [3.080, 3.120] GeV/$c^{2}$.

Potential remaining backgrounds are investigated by studying the inclusive $J/\psi$ and $\psi(3686)$ MC samples, using the event-type analysis tool TopoAna~\cite{TopoAna}. It is found that the peaking backgrounds are mainly from $J/\psi\to\pi^0\Sigma^{+}\overline{\Sigma}{}^-$ for the $J/\psi\to\eta\Sigma^{+}\overline{\Sigma}{}^-$ channel, and $\psi(3686)\to\gamma\chi_{c0,1,2},~\chi_{c0,1,2}\to\pi^{0}\Sigma^{+}\overline{\Sigma}{}^-$ for the $\psi(3686)\to\eta\Sigma^{+}\overline{\Sigma}{}^-$ channel.

\begin{figure}[htbp]
\centering
\begin{overpic}[width=7.0cm,height=5.0cm,angle=0]{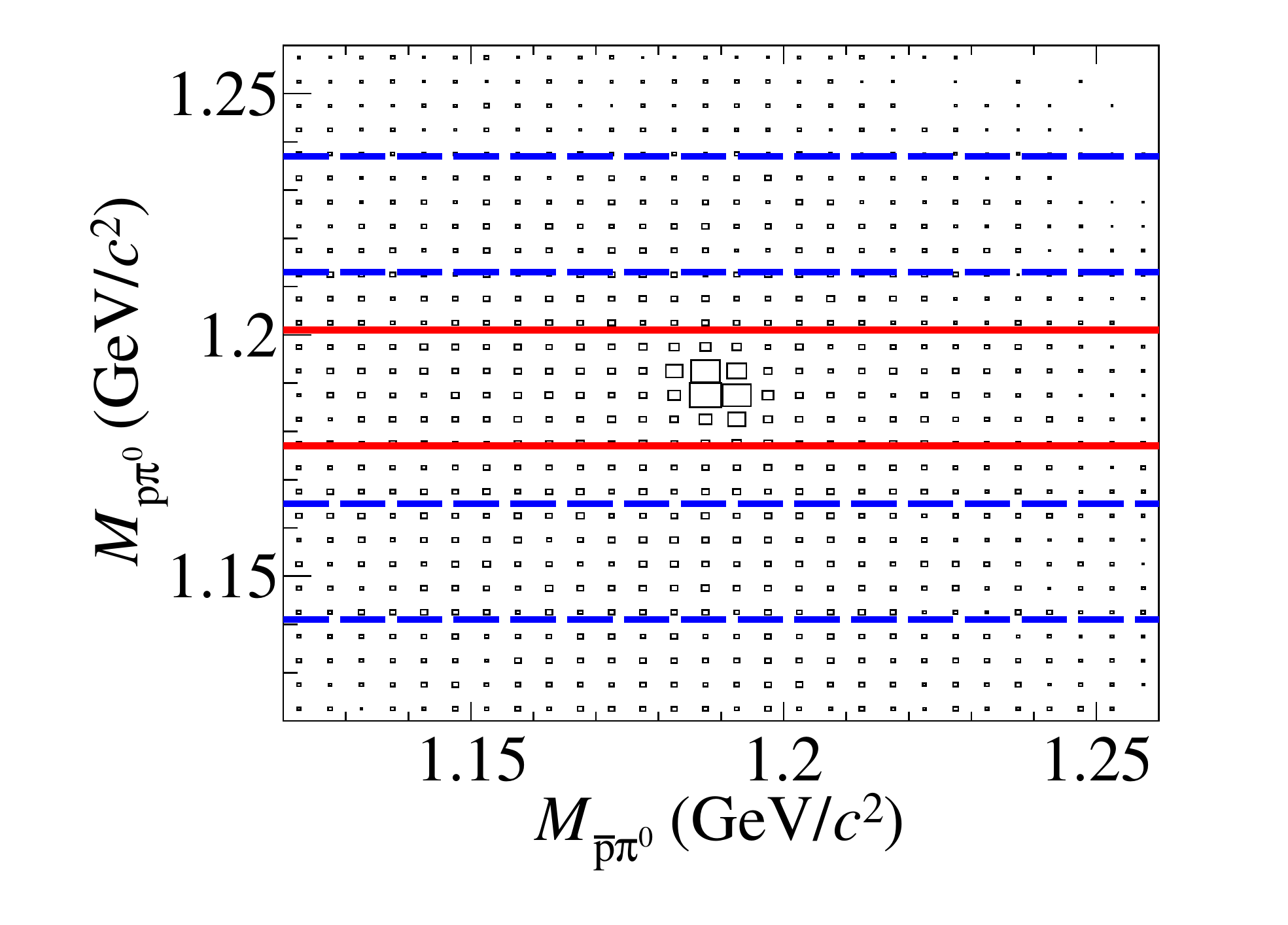}
\put(30,62){\normalsize\bf (a)}
\end{overpic}
\begin{overpic}[width=7.0cm,height=5.0cm,angle=0]{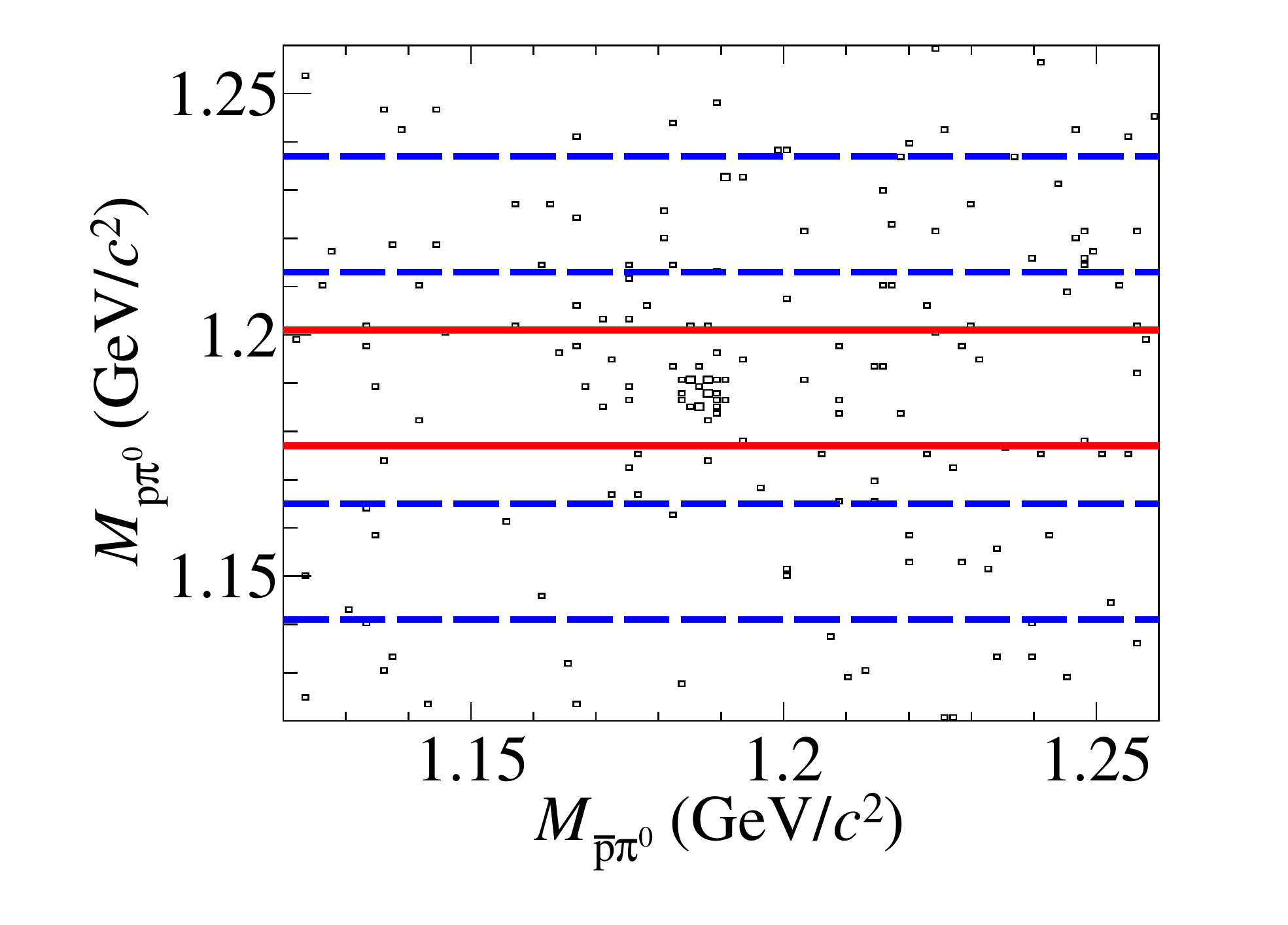}
\put(30,62){\normalsize\bf (b)}
\end{overpic}
\vskip -0.0cm
\parbox[1cm]{8cm} {
  \caption{The 2D distributions of $M_{p\pi^0}$ versus $M_{\bar{p}\pi^0}$ of the accepted candidates for (a) $J/\psi\to\eta\Sigma^{+}\overline{\Sigma}{}^-$ and (b) $\psi(3686)\to\eta\Sigma^{+}\overline{\Sigma}{}^-$, where the red solid lines and the blue dashed lines denote the $\Sigma$ signal and sideband regions, respectively.
}
\label{box}
}
\end{figure}

After imposing all the selection criteria, the two-dimensional (2D) distributions of the invariant mass of $p\pi^0$ ($M_{p\pi^0}$) versus the invariant mass of $\bar{p}\pi^0$ ($M_{\bar{p}\pi^0}$) of the accepted candidates for $J/\psi\to\eta\Sigma^{+}\overline{\Sigma}{}^-$ and $\psi(3686)\to\eta\Sigma^{+}\overline{\Sigma}{}^-$ in data are shown in Fig.~\ref{box}. A clear enhancement around the masses of $\Sigma^{+}$ and $\overline{\Sigma}^{-}$ is visible.
The $\Sigma$ signal and sideband regions are set to be $M_{p\pi^0}\in$ [1.177, 1.201] GeV/$c^{2}$ and $M_{p\pi^0}\in$[1.141, 1.165] GeV/$c^{2}$ or $M_{p\pi^0}\in$ [1.213, 1.237] GeV/$c^{2}$, respectively. Figure~\ref{fitresult} shows the $M_{\bar{p}\pi^0}$ distributions after requiring $M_{p\pi^0}$ to be within the $\Sigma^+$ signal region. A clear peak in the $\overline{\Sigma}^{-}$ region is visible.

The quantum electrodynamics (QED) production of $e^+e^-\to\eta\Sigma^{+}\overline{\Sigma}{}^-$ is studied using the off-resonance data taken at $\sqrt{s}=$ 3.080, 3.650 and 3.682 GeV. In the analysis no event satisfies the above selection criteria, thereby indicating that the background from the QED process is negligible.

\section{\boldmath Determination of the branching fractions}
\begin{figure}[htbp]
\centering
\begin{overpic}[width=7.0cm,height=5.0cm,angle=0]{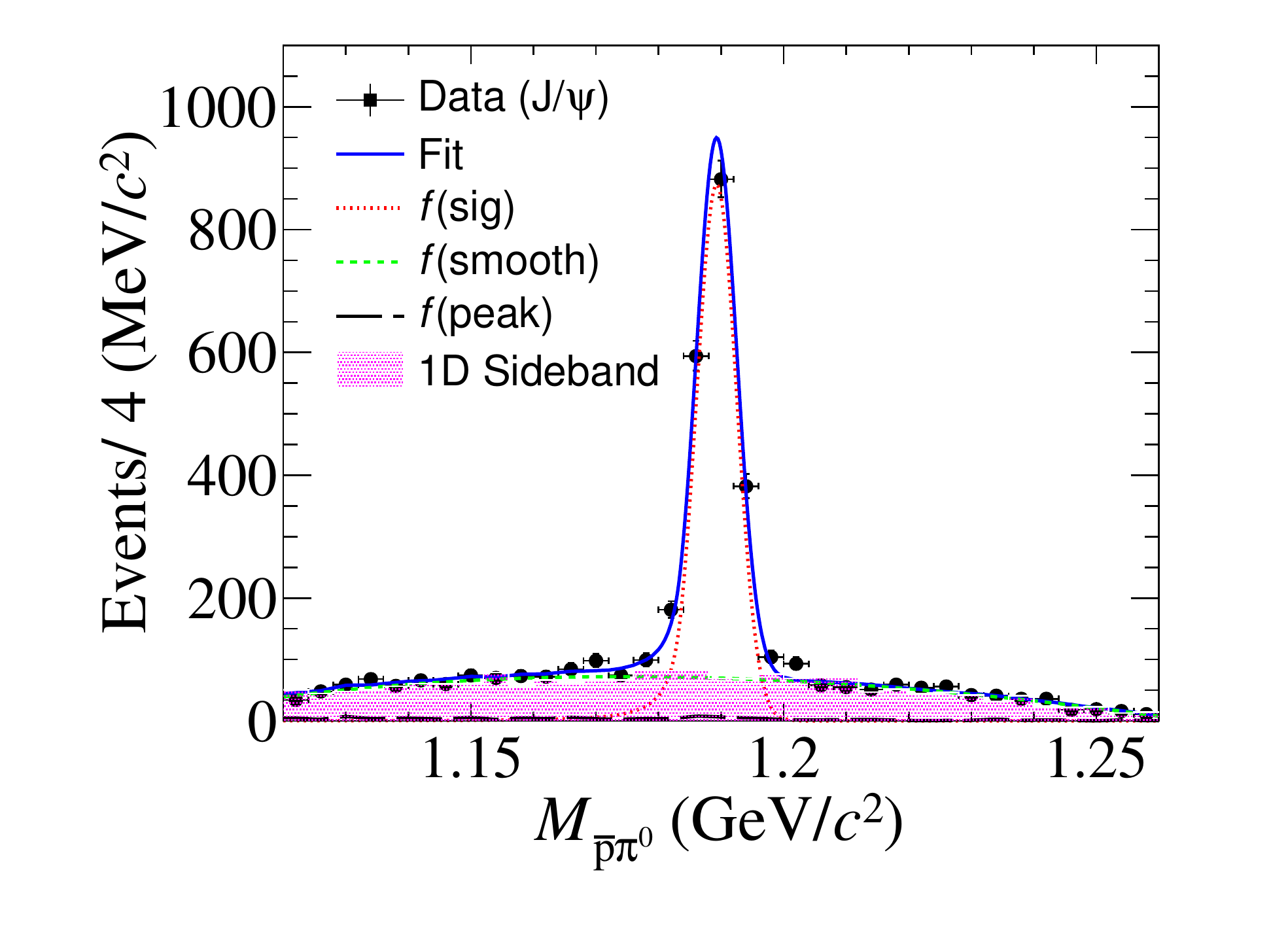}
\put(70,55){\normalsize\bf (a)}
\end{overpic}
\begin{overpic}[width=7.0cm,height=5.0cm,angle=0]{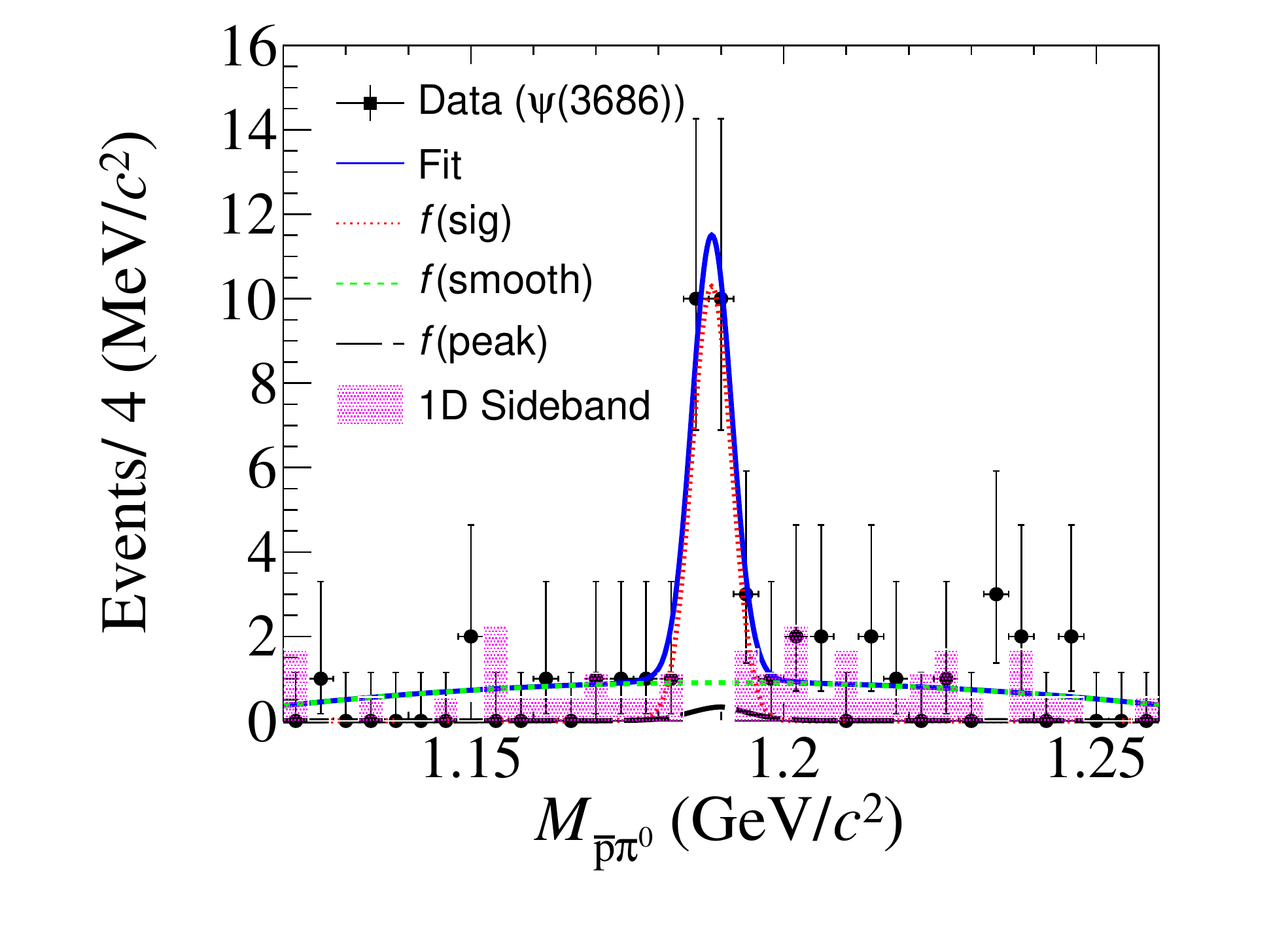}
\put(70,55){\normalsize\bf (b)}
\end{overpic}
\vskip -0.0cm
\parbox[1cm]{8cm} {
  \caption{Fits to the $M_{\bar{p}\pi^0}$ distributions of the accepted candidates for (a) $J/\psi\to\eta\Sigma^{+}\overline{\Sigma}{}^-$ and (b) $\psi(3686)\to\eta\Sigma^{+}\overline{\Sigma}{}^-$. The black points with uncertainties are data, the blue solid curves are the fit results, the red dotted lines denote the signal MC sample, the green dashed lines denote the Chebyshev function, the black long-dashed lines denote the backgrounds of $J/\psi\to\pi^0\Sigma^{+}\overline{\Sigma}{}^-$ ($J/\psi$ data) and $\psi(3686)\to\gamma\chi_{c0,1,2},~\chi_{0,1,2}\to\pi^0\Sigma^{+}\overline{\Sigma}{}^-$ ($\psi(3686)$ data). The pink shadow denote the scaled 1D sideband contribution according to the final fit results.
}
\label{fitresult}
}
\end{figure}

For the branching fraction measurement, the signal yield $N_{\rm obs}$ of the $\overline{\Sigma}^{-}$ peak is determined by an unbinned maximum likelihood fit to the $M_{\bar{p}\pi^0}$ distribution, as shown in Fig.~\ref{fitresult}. The $\overline{\Sigma}^{-}$ signal shape is described by a normalized Crystal Ball function~\cite{CBfunction}, since the distribution of the photon energy deposited in the EMC has a long tail on the low energy side. The smooth background shape is described by third-order and second-order Chebyshev functions for $J/\psi$ and $\psi(3686)$ decays, respectively, whose parameters are fixed from the fits to the sideband events and contributions are floated. The contribution of peaking backgrounds is described by the MC-simulated shapes obtained from the exclusive MC samples. To determine the expected yield of the peaking background, the control samples of $J/\psi\to\pi^0\Sigma^{+}\overline{\Sigma}{}^-$ and $\psi(3686)\to\gamma\chi_{c0,1,2},~\chi_{c0,1,2}\to\pi^{0}\Sigma^{+}\overline{\Sigma}{}^-$ are used. Based on the branching fractions obtained from the control samples and the detection efficiencies determined from the exclusive MC samples, we determined the yields of the peaking background to be $107.6 \pm 0.6$ and $1.1 \pm 0.1$ for $J/\psi$ and $\psi(3686)\to\eta\Sigma^{+}\overline{\Sigma}{}^-$, respectively. The statistical significance is estimated by the likelihood difference between the fits with and without the signal component, taking into account the modified number of the degrees of freedom. The fit is also performed by changing the fit range, the signal shape, or the background shape. In all cases, the statistical significance for $\psi(3686)\to\eta\Sigma^{+}\overline{\Sigma}{}^-$ and $J/\psi\to\eta\Sigma^{+}\overline{\Sigma}{}^-$ is greater than $5\sigma$. The signal yields are determined to be $1821.17 \pm 60.75$ and $20.49 \pm 5.07$ for $J/\psi$ and $\psi(3686)\to\eta\Sigma^{+}\overline{\Sigma}{}^-$, respectively, where the uncertainties are statistical only.

\begin{figure*}[htbp]
\centering
\begin{overpic}[width=7.0cm,height=5.0cm,angle=0]{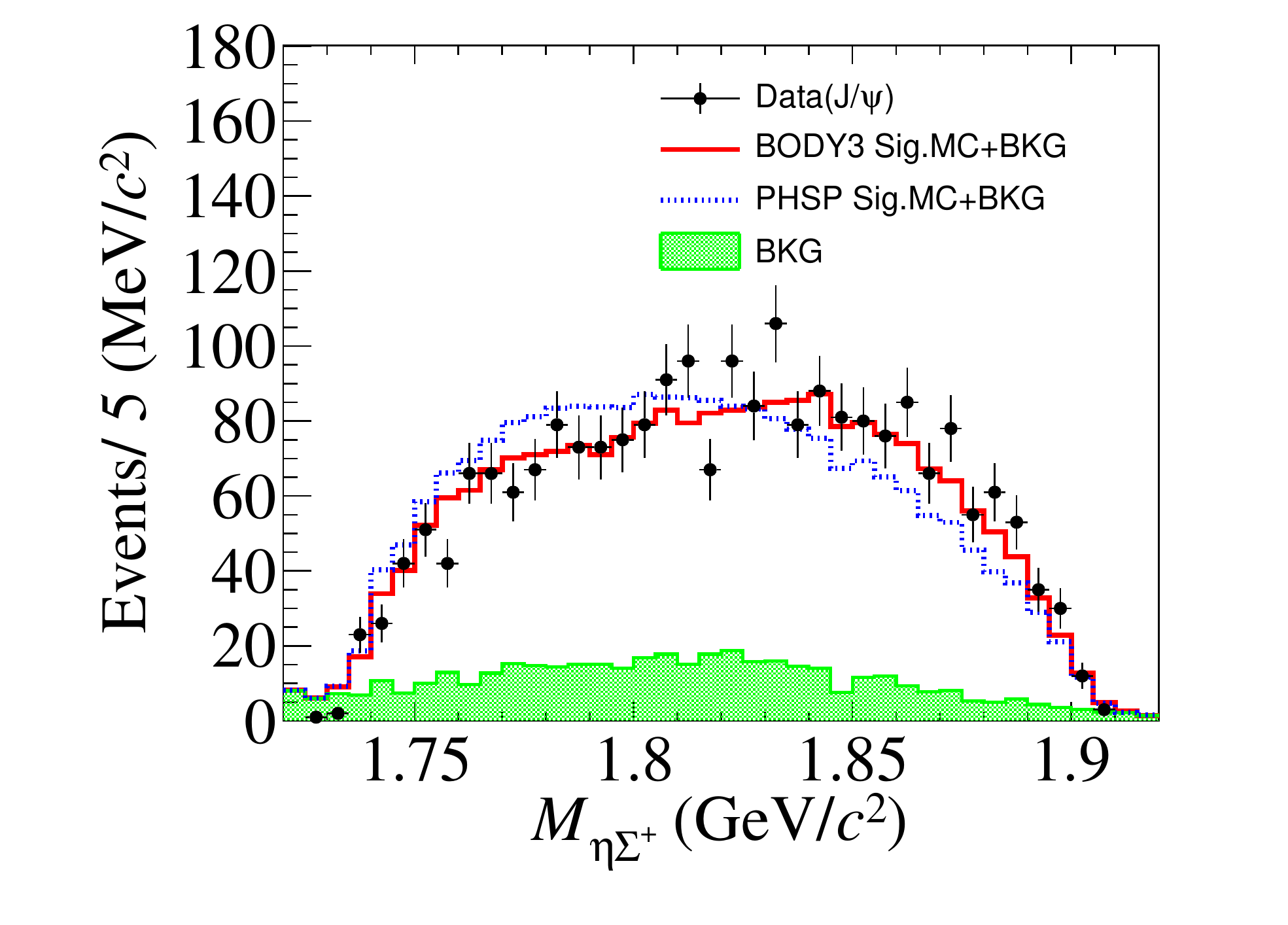}
\put(30,55){\normalsize\bf (a)}
\end{overpic}
\begin{overpic}[width=7.0cm,height=5.0cm,angle=0]{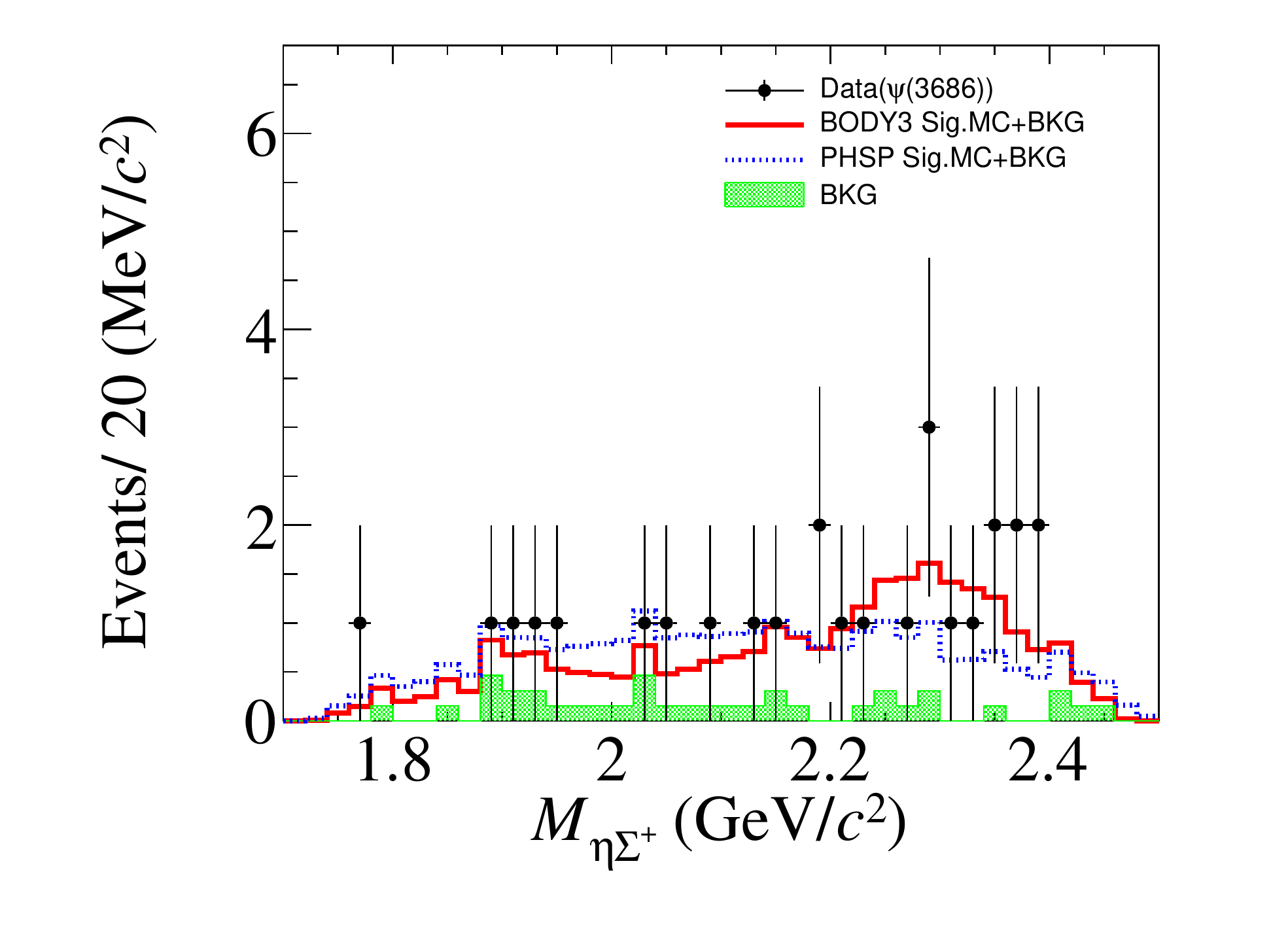}
\put(30,55){\normalsize\bf (d)}
\end{overpic}
\begin{overpic}[width=7.0cm,height=5.0cm,angle=0]{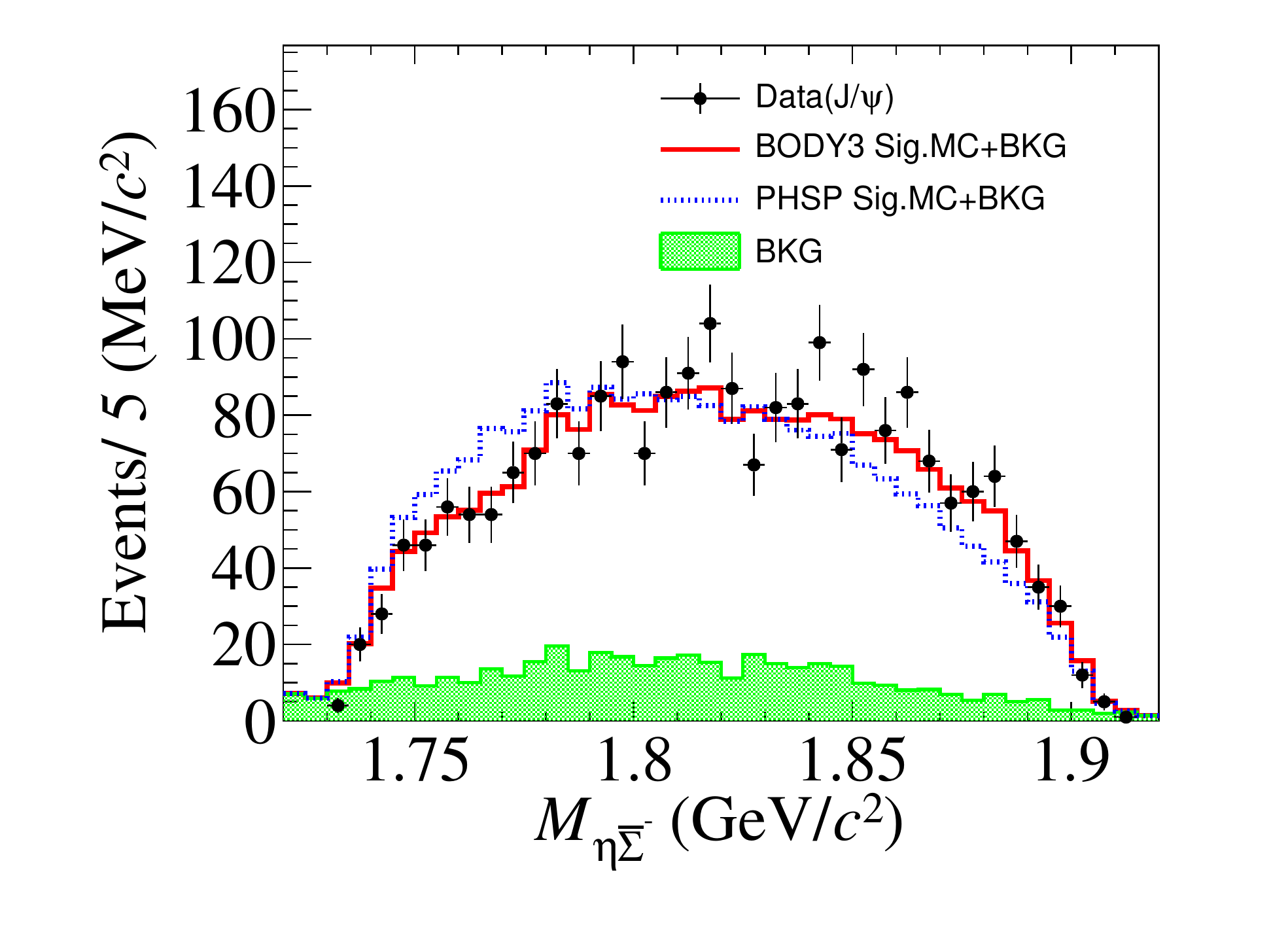}
\put(30,55){\normalsize\bf (b)}
\end{overpic}
\begin{overpic}[width=7.0cm,height=5.0cm,angle=0]{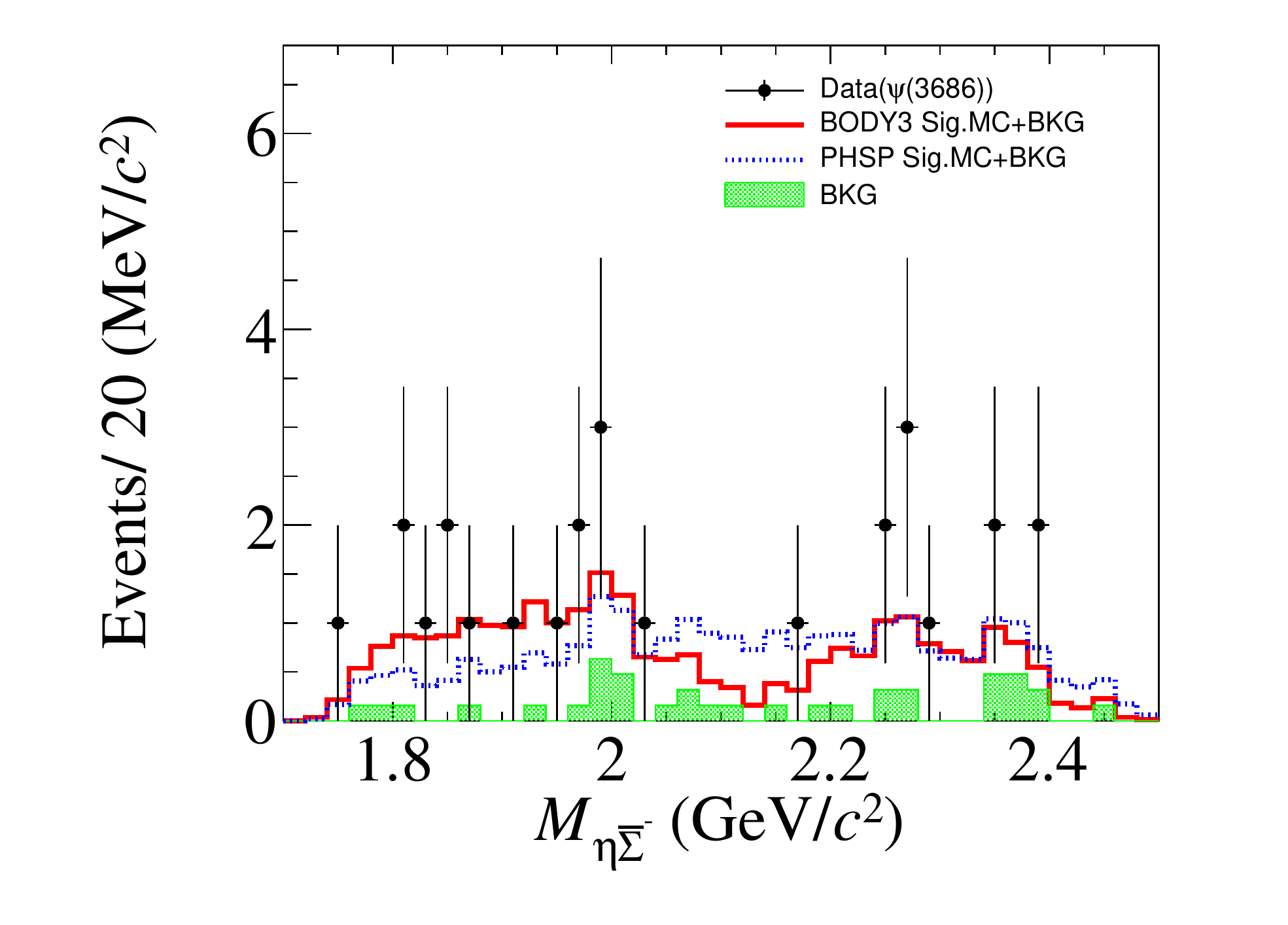}
\put(30,55){\normalsize\bf (e)}
\end{overpic}
\begin{overpic}[width=7.0cm,height=5.0cm,angle=0]{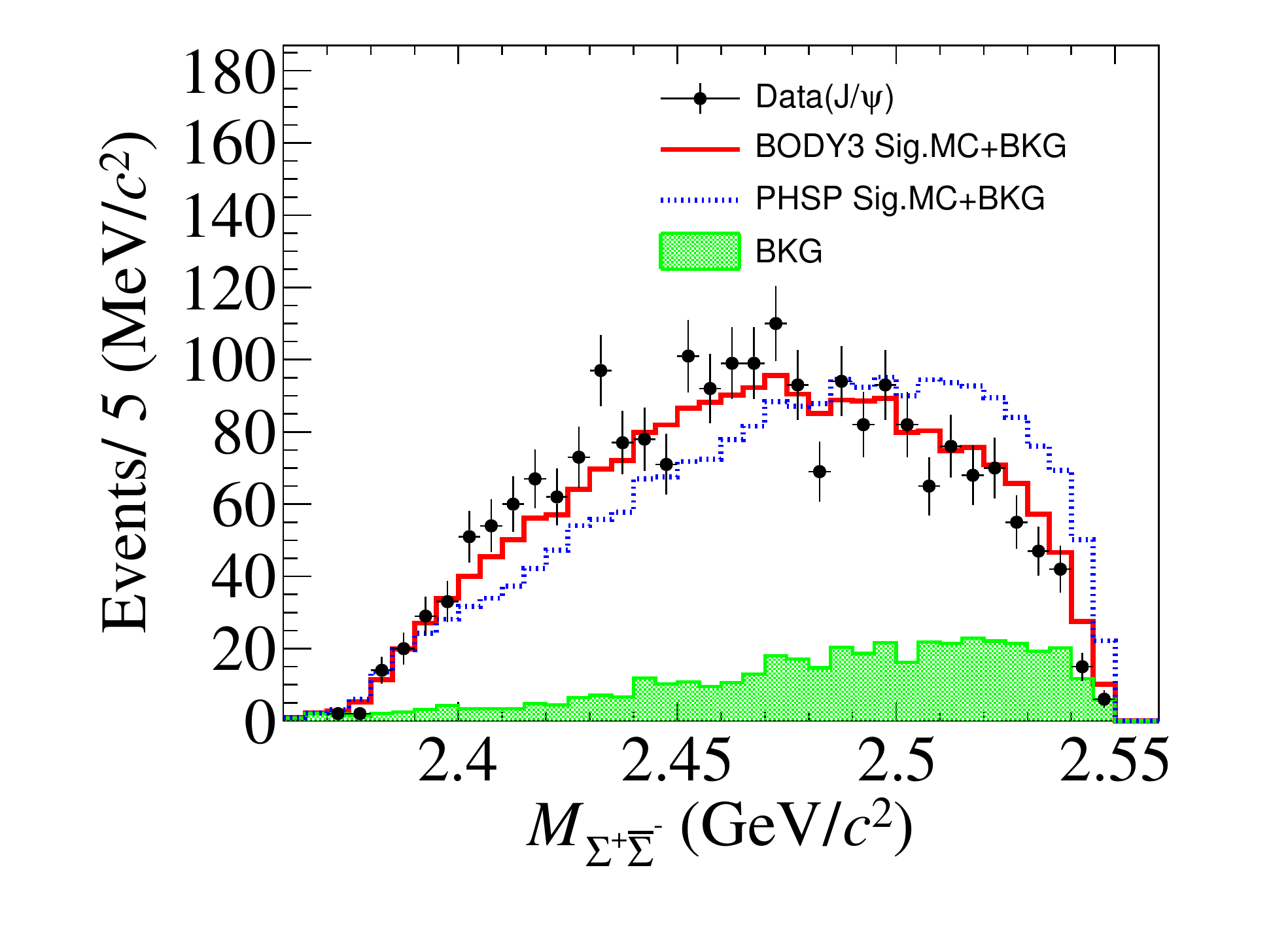}
\put(30,55){\normalsize\bf (c)}
\end{overpic}
\begin{overpic}[width=7.0cm,height=5.0cm,angle=0]{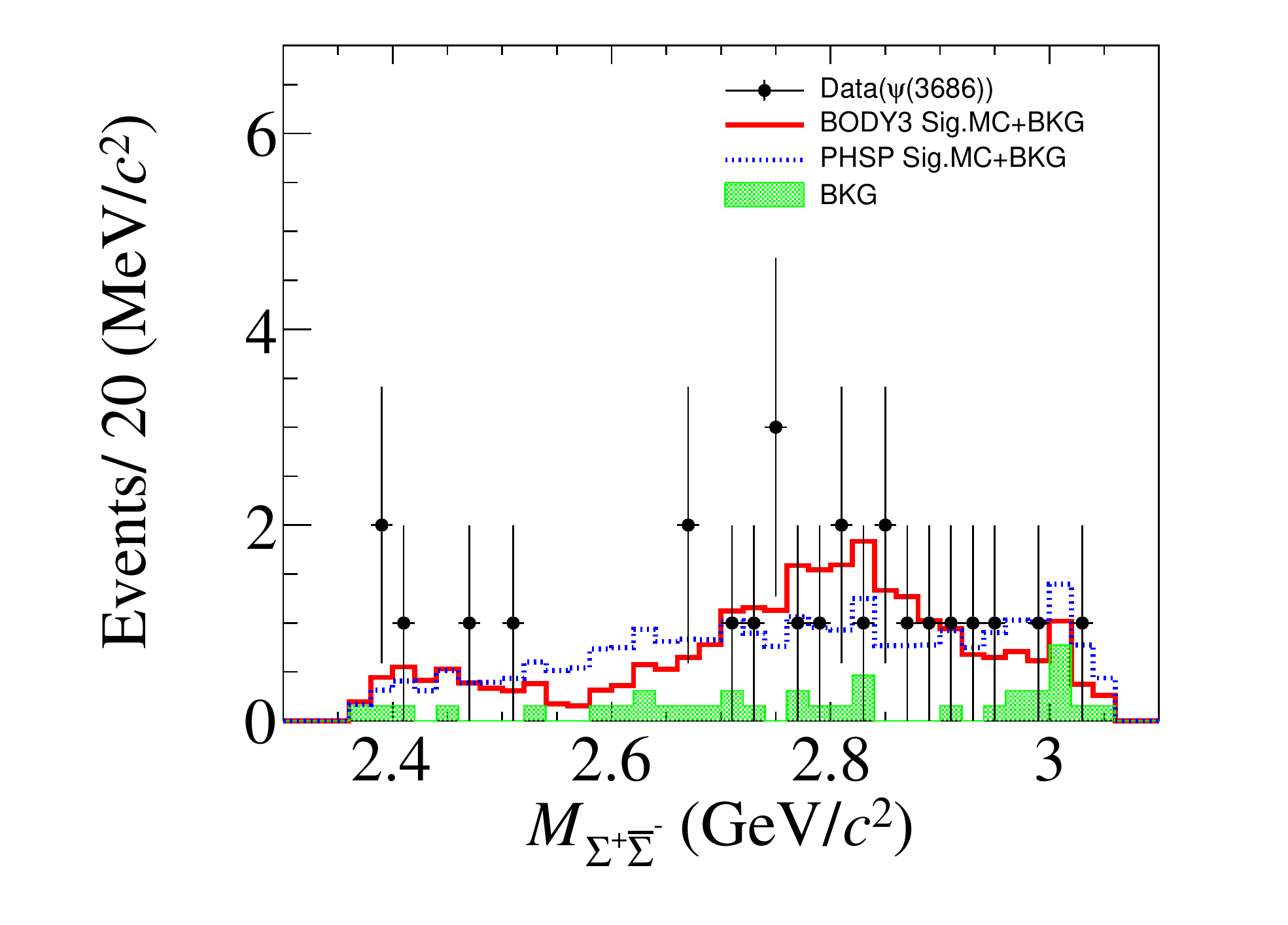}
\put(30,55){\normalsize\bf (f)}
\end{overpic}
\vskip -0.0cm
\parbox[1cm]{15cm}{
  \caption{Invariant mass distributions of all the two-body particle combinations for (left side) $J/\psi\to\eta\Sigma^{+}\overline{\Sigma}{}^-$ and (right side) $\psi(3686)\to\eta\Sigma^{+}\overline{\Sigma}{}^-$. The points with error bars are data, the red histograms are the sum of the sidebands and the signal MC sample generated with the modified data-driven generator, the blue dotted histograms are the sum of the sidebands and the signal MC sample generated with PHSP model, the green shaded histograms are the background contributions estimated from the $\Sigma$ sidebands. The signal and background yields have been normalized according to the fitting results for data.}
\label{project}
}
\end{figure*}

Figures~\ref{project} (a)-\ref{project}(f) show the invariant mass distributions of the different two-body particle combinations for $\psi\to\eta\Sigma^{+}\overline{\Sigma}{}^-$, where the background contributions are estimated from the $\Sigma$ sidebands. The experimental distributions deviate from the signal MC sample generated according to the phase space distribution (PHSP). To improve the reliability of the reconstruction efficiency $\varepsilon_{J/\psi\to\eta\Sigma^{+}\overline{\Sigma}{}^-}$, the PHSP model is replaced by the modified data-driven generator BODY3~\cite{BESIII7}, where the MC-simulated events are sampled according to the Dalitz distribution of the data to describe the potential intermediate states for a given three-body final state, obtaining good consistency. As shown in Fig.~\ref{project}, there is no structure visible in the $\eta\Sigma^{+}$, $\eta\overline{\Sigma}^{-}$ and $\Sigma^{+}\overline{\Sigma}{}^-$ invariant mass spectra.

The branching fraction of $\psi\to\eta\Sigma^{+}\overline{\Sigma}{}^-$ is calculated by
\begin{equation} \label{eq1}
{\cal B}(\psi\to\eta\Sigma^{+}\overline{\Sigma}{}^-) =\frac{N_{\rm obs} }{N^{\rm tot}_{\psi}\cdot\prod{\cal B}_{i}\cdot\epsilon},
\end{equation}
where $N_{\rm obs}$ is the number of signal events determined by the fit, $N^{\rm tot}_{\psi}$ is the number of the total $J/\psi$ or $\psi(3686)$ events~\cite{Njpsi, Npsip}, ${\cal B}_{i}$ is the branching fraction of the $i$th intermediate state taken from the PDG~\cite{PDG}, i.e. ${\cal B}(\pi^0\to\gamma\gamma)=(98.823\pm0.034)\%$, ${\cal B}(\eta\to\gamma\gamma)=(39.36\pm0.18)\%$ and ${\cal B}(\Sigma^{+}(\overline{\Sigma}^{-})\to\pi^0 p(\bar{p}))=(51.57\pm0.30)\%$, $\epsilon$ is the reconstruction efficiency, which is determined by the MC simulation based the BODY3 generator. The corresponding numerical values are listed in Table~\ref{calbr}.

\begin{table*}[htbp]
\caption{Summary of the number of $\psi$ events, the branching fractions of the intermediate states taken from the PDG~\cite{PDG}, the reconstruction efficiency, the correction factors and the signal yields used for branching fraction calculations. The uncertainties are statistical only.}
\begin{tabular}{ccc}
\hline\hline                                                &$J/\psi$ decay         &$\psi(3686)$ decay  \\\hline
$N^{\rm tot}_{J/\psi(\psi(3686))}$($\times 10^{6}$)           &10087 $\pm$ 44        &448.1 $\pm$ 2.9    \\
${\cal B}(\pi^0\to\gamma\gamma)$          & \multicolumn{2}{c}{$(98.823\pm0.034)$\%} \\
${\cal B}(\eta\to\gamma\gamma)$           & \multicolumn{2}{c}{$(39.36\pm0.18)$\%}\\
${\cal B}(\Sigma^{+}(\overline{\Sigma}^{-})\to\pi^0 p(\bar{p}))$     & \multicolumn{2}{c}{$(51.57\pm0.30)$\%}\\
Efficiency (\%)                                  &2.78 $\pm$ 0.01       &4.66 $\pm$ 0.01       \\
$N_{\rm obs}$                                   &1821.17 $\pm$ 60.75         &20.49 $\pm$ 5.07       \\\hline
${\cal B}$                                           &$\reJns$              &$\rePns$     \\
\hline
\hline
\end{tabular}
\label{calbr}
\end{table*}

\begin{table*}[htbp]
\caption{Relative systematic uncertainties (in \%) of the measurements of the branching fractions.}
\begin{tabular}{ccc}
\hline\hline
Source                                 &$J/\psi\to\eta\Sigma^{+}\overline{\Sigma}{}^-$          &$\psi(3686)\to\eta\Sigma^{+}\overline{\Sigma}{}^-$  \\\hline
Track reconstruction                   &3.0                        & ~3.0               \\
PID                                    &2.9                        & ~2.9               \\
$\pi^0$ reconstruction                 &1.4                        & ~1.0               \\
$\eta$ reconstruction                  &0.9                        & ~1.1               \\
Fit range                              &1.4                        & ~1.1               \\
Signal shape                           &1.9                        & ~0.5               \\
Smooth background                      &1.3                        & ~1.1               \\
Peaking background                     &0.6                        & ~1.0               \\
$p\bar{p}\eta'$ background veto	       &0.9	       		   & ~ -   		   \\
$\pi^0\pi^0J/\psi$ background veto      & -	                        & ~0.7   		   \\
Signal MC model                           &0.3                        & ~1.1                \\
Kinematic fit                          &2.0                        & ~3.6                \\
${\cal B}$ of intermediate state            &1.3                        & ~1.3               \\
$N_{\psi}^{\rm tot}$                   &0.4                        & ~0.6               \\\hline
Total                                  &5.9                        & ~6.3               \\
\hline
\hline
\end{tabular}
\label{sysall}
\end{table*}

\section{\boldmath Systematic uncertainty}
Several sources of systematic uncertainties for the branching fraction measurements are considered: the differences between data and MC simulation for track reconstruction, PID and $\pi^0$($\eta$) reconstruction, the uncertainty of the fitting model, the background substraction and description, the signal modeling, kinematic fit, the branching fractions of intermediate states, and the total number of $\psi$ events.

The uncertainties of track reconstruction efficiencies are estimated with the control sample $\psi(3686)\to p\bar{p}\pi^+\pi^-$~\cite{systrk}, and are determined to be 1.3\% and 1.7\% for each proton and antiproton, respectively. With the same control sample, the PID uncertainties are determined to be 1.3\% per proton and 1.6\% per antiproton.

The systematic uncertainty due to the $\pi^0$($\eta$) reconstruction efficiency is determined by using the control sample of $J/\psi\to p\bar{p}\pi^{0}(\eta)$ decays. The resulting systematic uncertainties of the $\pi^{0}$ reconstruction efficiency are determined to be 0.7\% and 0.5\% for the $J/\psi$ and $\psi(3686)$ decays, respectively, depending on the different $\pi^{0}$ momentum. The resulting systematic uncertainties of the $\eta$ reconstruction efficiency are determined to be 0.9\% and 1.1\% for the $J/\psi$ and $\psi(3686)$ decays, respectively, depending on the different $\eta$ momentum.

The systematic uncertainty of the fitting model originates from the fit range and the choice of the signal and the background functions. The uncertainty due to the fit range is estimated by varying the range by $\pm 10$ MeV/$c^{2}$. The largest difference of the resulting branching fractions is taken as the systematic uncertainty, which is 1.4\% and 1.1\% for the $J/\psi$ and $\psi(3686)$ decays, respectively. To estimate the uncertainties of the signal shape, a Breit-Wigner function convolved with a Gaussian function is used to replace the signal shape instead of the Crystal Ball function, while the background contributions are fixed to the nominal fit result. The differences to the nominal models, 1.9\% and 0.5\%, are taken as the systematic uncertainties for the $J/\psi$ and $\psi(3686)$ decays, respectively.
For the smooth background, the uncertainties are estimated by varying the order of the Chebychev polynomial function by $\pm 1$ order. The largest difference to the original function is taken as the systematic uncertainty, which is 1.3\% and 1.1\% for $J/\psi$ and $\psi(3686)$ decays, respectively.
For the peaking background, the systematic uncertainty for $J/\psi\to\pi^0\Sigma^{+}\overline{\Sigma}{}^-$ is estimated by removing and adding the background contribution in extracting the signal yield. The difference in the branching fraction determination, 0.6\%, is taken as the systematic uncertainty.
The systematic uncertainty for the background of $\psi(3686)\to\gamma\chi_{c0,1,2},~\chi_{c0,1,2}\to\pi^0\Sigma^{+}\overline{\Sigma}{}^-$ is estimated by changing the expected yield for peaking background events by $\pm 1\sigma$, where $\sigma$ is the uncertainty of $N_{\rm peak}$ mentioned above. The larger difference to the nominal result, 1.0\%, is taken as the systematic uncertainty.

To estimate the systematic uncertainty of the background veto for $J/\psi\to\eta\Sigma^{+}\overline{\Sigma}{}^-$, the background contribution $J/\psi\to\eta' p\bar{p},~\eta'\to\pi^0\pi^0\eta$ is subtracted by requiring the invariant mass of the $\pi^0\pi^0\eta$ combination outside the $\eta'$ signal window [0.95, 0.97] GeV/$c^2$. The associated systematic uncertainty is estimated by changing the $\eta'$ signal window by $\pm 1\sigma$, where the $\sigma$ denotes the mass resolution of $\eta'$. The largest change to the nominal result, 0.9\%, is taken as the systematic uncertainty.

For $\psi(3686)\to\eta\Sigma^{+}\overline{\Sigma}{}^-$, the systematic uncertainty of the requirement on the $M^{\rm rec}_{\pi^0\pi^0}$ is estimated by using the control sample of $\psi(3686)\to\pi^0\pi^0J/\psi,~J/\psi\to\eta p\bar{p}$. The efficiency, defined as the ratio of the number of signal events with and without the $M^{\rm rec}_{\pi^0\pi^0}$ requirement, is calculated and the difference between data and MC simulation values, 0.7\%, is taken as the systematic uncertainty.

The uncertainty due to the $M^{\rm rec}_{\eta}$ veto is ignored since the efficiency loss due to this requirement is negligible.

The systematic uncertainty of the signal MC modeling is estimated by varying the bin size of the input Dalitz plot by $\pm$10\%, and varying the background level in the input Dalitz plot in the BODY3 generator by $\pm 1\sigma$, where the $\sigma$ denotes the statistical uncertainty of the background level which is determined from the fit result. Combining the results from the two sources, the largest change to the nominal reconstruction efficiency, 0.3\% and 1.1\%, are taken as the systematic uncertainties for the $J/\psi$ and $\psi(3686)$ decays, respectively.

The systematic uncertainty of the kinematic fit is estimated by using the control sample of $\psi(3686)\to\pi^0\pi^0J/\psi,~J/\psi\to p\bar{p}\eta$. The efficiency of kinematic fit is defined as the ratio of the number of signal events with and without the kinematic fit. The differences of the efficiencies between data and MC simulation are determined to be 2.0\% and 3.6\% for $J/\psi$ and $\psi(3686)$ decays, respectively, depending on the different $\chi^{2}_{\rm 7C}$ requirement.

The uncertainties from the quoted branching fractions of $\eta\to\gamma\gamma$, $\Sigma^{+}(\overline{\Sigma}^{-})\to p(\bar{p})\pi^{0}$, $\pi^0\to\gamma\gamma$~\cite{PDG} are 0.5\%, 0.6\% and less than 0.1\%, respectively, and the total uncertainty is determined to be 1.3\%.

The systematic uncertainty from the total number of $\psi$ events, which are determined with inclusive hadronic events, are 0.4\% and 0.6\% for $J/\psi$ and $\psi(3686)$ data samples, respectively~\cite{Njpsi, Npsip}.

Table~\ref{sysall} lists all the systematic uncertainty contributions on the branching fraction measurements. The total systematic uncertainty is obtained by adding the individual contributions in quadrature. The total systematic uncertainties are 5.9\% and 6.3\% for $J/\psi\to\eta\Sigma^{+}\overline{\Sigma}{}^-$ and $\psi(3686)\to\eta\Sigma^{+}\overline{\Sigma}{}^-$, respectively.

\section{\boldmath Summary and discussion}
Using the data samples of $(10087 \pm 44)\times 10^{6}$ $J/\psi$ and $(448.1 \pm 2.9)\times 10^{6}$ $\psi(3686)$ events collected with the BESIII detector, the decays $J/\psi\to\eta\Sigma^{+}\overline{\Sigma}{}^-$ and $\psi(3686)\to\eta\Sigma^{+}\overline{\Sigma}{}^-$ are observed for the first time. The branching fractions of these two decays are determined to be ${\cal B}(J/\psi\to\eta\Sigma^{+}\overline{\Sigma}{}^-)=\reJ$ and ${\cal B}(\psi(3686)\to\eta\Sigma^{+}\overline{\Sigma}{}^-)=\reP$, where the first uncertainties are statistical and the second are systematic. The ratio of these two branching fractions is determined to be $\frac{{\cal B}(\psi(3686)\to\eta\Sigma^{+}\overline{\Sigma}{}^-)}{{\cal B}(J/\psi\to\eta\Sigma^{+}\overline{\Sigma}{}^-)} = (\reQ) \%$, where the uncertainty includes the statistical uncertainty and the uncorrelated systematic uncertainty, which is in agreement with the ``12\% rule".
No significant structures are observed in the $\eta\Sigma^{+}$, $\eta\overline{\Sigma}^{-}$, and $\Sigma^{+}\overline{\Sigma}{}^-$ invariant mass spectra.
However, the shapes of the invariant mass distributions of all subsystems deviate from the pure 3-body decay distribution. This implies the existence of some unknown dynamical effect. A partial wave analysis applied in a larger data sample may lead to decouple the underlying dynamics of the phenomenon~\cite{whitepaper}.

\acknowledgments
The BESIII Collaboration thanks the staff of BEPCII and the IHEP computing center for their strong support. This work is supported in part by National Key R$\&$D Program of China under Contracts No. 2020YFA0406300, No. 2020YFA0406400; National Natural Science Foundation of China (NSFC) under Contracts No. 11635010, No. 11735014, No. 11835012, No. 11935015, No. 11935016, No. 11935018, No. 11961141012, No. 12022510, No. 12025502, No. 12035009, No. 12035013, No. 12192260, No. 12192261, No. 12192262, No. 12192263, No. 12192264, No. 12192265; the Chinese Academy of Sciences (CAS) Large-Scale Scientific Facility Program; Joint Large-Scale Scientific Facility Funds of the NSFC and CAS under Contract No. U1832207; the CAS Center for Excellence in Particle Physice (CCEPP); 100 Talents Program of CAS; The Institute of Nuclear and Particle Physics (INPAC) and Shanghai Key Laboratory for Particle Physics and Cosmology; ERC under Contract No. 758462; European Union's Horizon 2020 research and innovation programme under Marie Sklodowska-Curie grant agreement under Contract No. 894790; German Research Foundation DFG under Contracts No. 443159800, No. 455635585, Collaborative Research Center CRC 1044, FOR5327, GRK 2149; Istituto Nazionale di Fisica Nucleare, Italy; Ministry of Development of Turkey under Contract No. DPT2006K-120470; National Science and Technology fund; National Science Research and Innovation Fund (NSRF) via the Program Management Unit for Human Resources $\&$ Institutional Development, Research and Innovation under Contract No. B16F640076; Olle Engkvist Foundation under Contract No. 200-0605; STFC (United Kingdom); Suranaree University of Technology (SUT), Thailand Science Research and Innovation (TSRI), and National Science Research and Innovation Fund (NSRF) under Contract No. 160355; The Royal Society, UK under Contracts No. DH140054, No. DH160214; The Swedish Research Council; U. S. Department of Energy under Contract No. DE-FG02-05ER41374.

\end{document}